\documentclass[]{aa}
\usepackage{graphics,epsfig,txfonts}
\usepackage{natbib}
\bibpunct{(}{)}{;}{a}{}{,}

\usepackage{graphicx}

\usepackage{longtable}
\usepackage{amssymb}
\usepackage{color}
\usepackage{lipsum}
\usepackage{textcomp}
\usepackage[colorlinks=true,linkcolor=blue,citecolor=blue,
bookmarks=true,bookmarksopen=true,bookmarksnumbered=true,draft=false]{hyperref}
\usepackage{enumerate}

\usepackage{pdflscape}

\newcommand{\oergs}[1]{$10^{#1}$ erg s$^{-1}$}

\newcommand{\ltsima}{$\buildrel < \over \sim$}
\newcommand{\lsim}{\lower.5ex\hbox{\ltsima}}
\newcommand{\gtsima}{$\buildrel > \over \sim$}
\newcommand{\gsim}{\lower.5ex\hbox{\gtsima}}

\newcommand{\xmm}{{\it XMM-Newton}\xspace}

\newcommand{\cxo}{{\it Chandra}\xspace}

\begin{document}

\title{Searching for intermediate-mass black holes in galaxies with low-luminosity AGN: a multiple-method approach }

\author{Filippos Koliopanos\inst{1,2}\thanks{\email{fkoliopanos@irap.omp.eu}}
  \and  Bogdan C. Ciambur\inst{3} 
  \and  Alister W. Graham\inst{3}
  \and  Natalie A. Webb\inst{1,2}
  \and  Mickael Coriat\inst{1,2}
  \and  Bur\c{c}in Mutlu-Pakdil\inst{4,5}
  \and  Benjamin L. Davis\inst{3}
  \and  Olivier Godet\inst{1,2}
  \and  Didier Barret\inst{1,2}
  \and  Marc S. Seigar\inst{4}}

\titlerunning{Searching for IMBHs in Low Luminosity AGN}
\authorrunning{Koliopanos et al.}

\institute{CNRS, IRAP, 9 Av. colonel Roche, BP 44346, F-31028 Toulouse cedex 4, France
          \and Universit{\'e} de Toulouse; UPS-OMP; IRAP, Toulouse, France 
          \and Centre for Astrophysics and Supercomputing, Swinburne University of Technology, Hawthorn, VIC 3122, Australia 
          \and Department of Physics and Astronomy, University of Minnesota Duluth, Duluth, MN 55812, USA
          \and Minnesota Institute for Astrophysics, University of Minnesota, Twin Cities, MN 55455, USA }

\date{Received 14 November 2016 / Accepted 20 December 2016}

  \abstract
   {}
   {This work is the first stage of a campaign to search for intermediate-mass black holes (IMBHs) in low-luminosity active Galactic nuclei (LLAGN) and dwarf galaxies. An additional and equally important aim of this pilot study is to investigate the consistency between the predictions of several popular black hole scaling relations and the fundamental plane (FP) of black-hole activity (FP-BH).}
     {We used well established X-ray and radio luminosity relations in accreting black holes, along with the latest scaling relations between the mass of the central black hole ($M_{\rm BH}$) and the properties of its host spheroid, to predict $M_{\rm BH}$ in seven LLAGN, that were previously reported to be in the IMBH regime. Namely, we used the recently re-evaluated $M_{\rm BH} - M_{\rm sph}$ ($M_{\rm sph}$: spheroid absolute magnitude at 3.6\,${\mu}{\rm m}$) scaling relation for spiral galaxies,  the $M_{\rm BH} - n_{\rm sph}$ ($n_{\rm sph}$:  major axis S\'ersic index of the spheroid component) relation, the $M_{\rm BH} -PA$ ($PA$:  spiral pitch angle) relation, and a recently re-calibrated version of the FP-BH for weakly accreting BHs, to independently estimate $M_{\rm BH}$ in all seven galaxies.}
   {We find that all LLAGN in our list have low-mass central black holes with $\log{M_{\rm BH}/M_{\odot}}{\approx}6.5$ on average,  but that they are, most likely, not IMBHs. All four methods used predicted consistent BH masses in the 1$\sigma$ range. Furthermore, we report that, in contrast to previous classification, galaxy NGC~4470 is bulge-less, and we also cast  doubts on the AGN classification of NGC~3507.}
   { We find that our latest, state-of-the-art techniques for bulge magnitude \& S\'ersic index computations and the most recent updates of the $M_{\rm BH} - L_{\rm sph}$, $M_{\rm BH} - n_{\rm sph}$, and $M_{\rm BH} -PA$ relations and the FP-BH produce consistent results in the low-mass regime. We thus establish a multiple-method approach for predicting BH masses in the regime where their spheres of gravitational influence cannot be spatially resolved. Our approach mitigates against outliers from any one relation and provides a more robust average prediction. 
   We will use our new method to revisit more IMBH candidates in LLAGN.}
   {}

\maketitle

\section{Introduction}
\label{sec-intro}
The existence of stellar-mass (${<}10^{2}\,{\rm M}_\odot$)  black holes (BHs) and supermassive (${>}10^{5}\,{\rm M}_\odot$) BHs (SMBHs) has been firmly established by numerous observations of BH X-ray binaries \citep[e.g.,][and references therein]{2006ARA&A..44...49R,2014SSRv..183..223C}, recent gravitational wave detections of BH mergers \citep{2016PhRvL.116x1103A,2016PhRvL.116f1102A}, and abundant observational evidence in favor of active galactic nuclei (AGN) being powered by SMBHs (evidence from X-ray emission, e.g., \citealt{1995Natur.375..659T}; \citealt{1997ApJ...488L..91N}; \citealt{2004ApJ...604...63Y} or H$_{2}$O maser emission, e.g., \citealt{1995Natur.373..127M}; \citealt{2003ApJ...582L..11G}), which are most likely present in the center of most massive galaxies (indications of weakly accreting SMBHs in most nearby galaxies, e.g., \citealt{1997ApJS..112..315H}; \citealt{2000MNRAS.315...98R}; \citealt{2011ApJS..192...10L} or direct mass measurements of the SMBH in the Milky Way and nearby galaxies e.g., \citealt{2002Natur.419..694S}; \citealt{2005ApJ...620..744G}; \citealt{2009ApJ...692.1075G} and \citealt{2001ApJ...555..685B}; \citealt{2011ApJ...727...20K}, respectively). Nevertheless, there is a conspicuous scarcity of BHs in the mass regime between these two boundaries (namely, $10^{2}-10^{5}\,{\rm M}_\odot$). The objects expected to occupy this gap in the BH mass range are known as intermediate-mass BHs (IMBHs). The search for IMBHs is important, not only because their existence is predicted by a variety of plausible scenarios (evolution of Population III stars: e.g., \citealt{2001ApJ...551L..27M}; \citealt{2002ApJ...571...30S}; \citealt{2016MNRAS.460.4122R}, repeated mergers onto massive stars in young massive clusters \citealt{2002ApJ...576..899P}; \citealt{2004Natur.428..724P}; \citealt{2004ApJ...604..632G}; \citealt{2006MNRAS.368..141F} or BH mergers in dense clusters: e.g., \citealt{1987ApJ...319..772L}; \citealt{1990ApJ...356..483Q}; \citealt{1995MNRAS.272..605L}; \citealt{2002ApJ...566L..17M}; \citealt{2002ApJ...576..894M}; \citealt{2002MNRAS.330..232C}), but also because they are proposed to be seeds for SMBHs \citep{2001ApJ...552..459H,2001ApJ...562L..19E,2004IJMPD..13....1M}. Discovering IMBHs and studying their radiative properties will also advance our understanding of the effects of heating and radiation pressure on their surrounding gas, ultimately elucidating the mechanisms of early galaxy formation \citep{2009ApJ...698..766M,2009ApJ...696L.146M,2016MNRAS.460.4122R}. At the same time, discovering active IMBHs in the center of low-mass galaxies, will further our insight into BH feedback and galaxy formation in different mass regimes \citep[][and references therein]{2013ApJ...775..116R,2016ApJ...818..172G}. 

One of the methods used in the search for IMBHs involves observing accreting BHs and exploiting the limits imposed by physical laws on their accretion rate (and therefore luminosity), that is, the Eddington limit. Namely, ultra-luminous ($L_{\rm x}{>}$\oergs{39}) X-ray sources (ULXs) had been initially  considered as IMBH candidates \citep[see review by][]{2011NewAR..55..166F}, however,  later studies \citep[e.g.,][and references therein]{2013MNRAS.435.1758S, 2013ApJ...778..163B} have demonstrated that super-Eddington accretion onto stellar BHs may account for most ULXs with luminosities of up to $\sim$\oergs{40}. Furthermore, \cite{2014Natur.514..202B} showed that luminosities of up to $\sim$\oergs{40} can be achieved by super-Eddington accretion onto a neutron star (NS). Nevertheless, there is a small group of hyper-luminous X-ray sources (HLXs) with luminosities exceeding $\sim$\oergs{41} that are hard to explain, even when invoking super-Eddington accretion and/or beaming of the X-ray emission. HLXs are strong IMBH candidates, with a particular source, known as ESO 243-49  HLX-1 (henceforth HLX-1), considered as the best IMBH candidate found so far.  Serendipitously discovered by \cite{2009Natur.460...73F} in the 2XMM catalog \citep{2009A&A...493..339W}, HLX-1 was apparently associated with the galaxy ESO 243-49 at a distance of 95\,Mpc. The optical counterpart located by \cite{2010MNRAS.405..870S}, using the excellent X-ray position derived from Chandra data \citep{2010ApJ...712L.107W}, shows an H$_\alpha$ emission line with a redshift similar to the redshift of ESO 243-49, thus confirming the association with ESO 243-49 \citep{2010ApJ...721L.102W}. Taking the maximum unabsorbed X-ray luminosity of 1.1 $\times$ 10$^{42}$ erg s$^{-1}$ (0.2-10.0\,keV) and assuming that this value exceeds the Eddington limit by at most a factor of ten, implies a minimum mass of 500 M$_\odot$ \citep{2009Natur.460...73F}, making HLX-1 a prime IMBH candidate.

In addition to the search for HLXs, IMBHs can, in principle, be discovered in the center of low-luminosity AGN (LLAGN), and ``dwarf'' galaxies. These low-mass galaxies are expected to have undergone quiet merger histories and are, therefore, more likely to host lower-mass central black holes, a fraction of which is expected to lie in the IMBH range. In contrast to HLXs, LLAGN are accreting at rates well below the Eddington limit. Nevertheless, the mass of their accreting BH can be measured using X-ray and radio observations and employing the ``fundamental plane of black hole activity'' \citep[FP-BH,][]{2003MNRAS.345.1057M, 2004A&A...414..895F}. Accretion disks in active BHs during episodes of low-luminosity advection-dominated accretion (also known as a {\it hard state}) are usually accompanied by a relativistic jet \citep[e.g.,][]{2003MNRAS.344...60G,2005Ap&SS.300..177N,2009MNRAS.396.1370F} and produce emission that ranges from the radio to the X-ray band. The radio emission is associated with the jet and is most likely the result of synchrotron emission \citep[e.g.,][]{1977MNRAS.179..433B,1982MNRAS.199..883B,1984RvMP...56..255B} and non-thermal X-ray radiation is expected to originate from a corona of hot thermal electrons in the vicinity of the compact object. However, the entire broadband emission can also be dominated by the jet itself, during this state, especially at relatively high accretion rates (${\sim}0.01\dot M_{\rm Edd}$, e.g. \citealt{2004A&A...414..895F}).   An additional, less prominent thermal component originating from the truncated accretion disk \citep{1973A&A....24..337S} may also be registered along with the primary hard X-ray emission. Since the production of the jet is thought to be linked directly to the accretion process, it can be shown that the luminosities of the radio and X-ray emission are correlated and the disk-jet mechanism is scale-invariant with respect to the BH mass \citep{2003MNRAS.343L..59H}. Based on these theoretical predictions, \cite{2003MNRAS.345.1057M} investigated a large sample of Galactic BHs and SMBHs and found a strong correlation between the radio luminosity ($ {\rm L_{r}}$, 5\,GHz), the X-ray luminosity ($ {\rm L_{x}}$, 0.5-10\,keV), and the BH mass (${\rm M_{\rm BH}}$), known as the FP-BH. In the case of HLX-1, \cite{2012Sci...337..554W} used this relation to place an upper limit of less than  $ 10^5\,M_\odot$ for the mass of the BH in the system.  This was in agreement with the estimation of a mass of $\sim$10$^4$ M$_\odot$ by \cite{2012ApJ...752...34G}, \cite{2011ApJ...734..111D}, and \cite{2014A&A...569A.116S} using accretion disk modeling.

Another approach in the search for IMBHs draws on the correlations between large-scale properties of galaxies and the mass of their central SMBH. These involve the well-known scaling relations between BH mass and stellar bulge velocity dispersion \citep[$M_{{\rm BH}}$--$\sigma$: e.g.,][]{2000ApJ...539L...9F,2000ApJ...539L..13G,2011MNRAS.412.2211G,2013ApJ...764..184M,2015MNRAS.446.2330S} and BH mass and bulge luminosity \citep[$M_{{\rm BH}}$--$L$: e.g.,][]{1989IAUS..134..217D, 1993nag..conf.....B,1998AJ....115.2285M,2003ApJ...589L..21M, 2013ApJ...764..151G,2016ApJ...817...21S}. 
Studying low-mass galaxies using the 
$M_{{\rm BH}}$--$\sigma$, 
$M_{{\rm BH}}$--$L$ relations may reveal new IMBH candidates. 
Indeed, applying the $M_{{\rm BH}}$--$\sigma$ relation to samples of low-mass galaxies has led to the identification of multiple (tentative) IMBH candidates \citep[e.g.,][]{2004ApJ...610..722G,2008AJ....136.1179B,2012ApJ...755..167D}. However, these samples, suffer from luminosity bias and large uncertainties that are partially due to the precarious application of the $M_{{\rm BH}}$--$\sigma$ relation in the low-mass regime \citep[e.g., see discussion in][]{2007ApJ...667..131G, 2014AJ....148..136M}. 
Additionally, the application of the $M_{{\rm BH}}$--$L$ relation in low-luminosity/low-mass galaxies to the search for IMBHs has also been problematic, since most galactic samples with directly measured SMBH masses, and thus the scaling relations, were dominated by luminous spheroids \citep[e.g.,][]{2003ApJ...589L..21M,2007AAS...211.1327G}. 
However, \cite{2012ApJ...746..113G} dramatically revised the well-known, near-linear relationship relating the SMBH mass and the host galaxy spheroid mass, demonstrating that it is  approximately quadratic for the low- and intermediate-mass spheroids. \cite{2013ApJ...764..151G} (henceforth GS13) and \cite{2013ApJ...768...76S} built on the new $M_{{\rm BH}}$--spheroid mass relation, and by including many more low-mass spheroids with directly measured SMBH masses, they showed that the $M_{{\rm BH}}$--$L$ relation is better described by two power-laws \citep[see also][]{2015ApJ...798...54G}. Using this improved scaling relation, GS13 identified 40 lower luminosity spheroids that contain AGN according to the spheroid magnitudes reported by \cite{2006AJ....131.1236D} (henceforth D\&D06) and appear to have a mass of the central black hole that falls in the IMBH range. Recently, the scaling relation of GS13 was refined by \cite{2016ApJ...817...21S} using a sample of 66 nearby galaxies with a dynamically measured BH mass.

\cite{2007ApJ...655...77G}, \cite{2013MNRAS.434..387S}, and \cite{2016ApJ...817...21S} revisited the previously reported \citep{2001ApJ...563L..11G,2003RMxAC..17..196G} correlation between the mass of a galaxy's central BH and its bulge concentration and constructed a relation between $M_{\rm BH}$ and the galactic S\'ersic index  (which is a measure of bulge concentration). Furthermore, \cite{2008ApJ...678L..93S} discovered a correlation between the morphology of the spiral arms and the mass of the central BH in disk galaxies \citep[see also][]{2012ApJS..199...33D}. This, was formulated as a relation between $M_{\rm BH}$ and the spiral pitch angle (a measure of the tightness of spiral arms). These methods benefit from the fact that they do not require spectroscopy and are independent of the source distance.

Motivated by the promising findings of GS13, we are initiating an extensive search for IMBH candidates in LLAGN using a multiple method approach. Namely, we use radio, infrared (IR), and X-ray observations of low luminosity spheroids with AGN and employ the FP-BH, together with the most up to date BH scaling relations with spheroid properties and with spiral arm pitch angle (henceforth $PA$), to determine the mass of the central BH. More specifically, we make use of the FP-BH re-calibrated by \cite{2012MNRAS.419..267P} to account for weakly accreting, low-luminosity BHs, the latest $M_{\rm BH} - M_{\rm sph}$ ($M_{\rm sph}$: spheroid absolute magnitude at 3.6\,${\mu}{\rm m}$) scaling relation for spiral galaxies by \citep{2016ApJ...817...21S},  the $M_{\rm BH} - n_{\rm sph}$ ($n_{\rm sph}$:  major axis S\'ersic indices of the spheroid components) relation reported in  \cite{2016ApJ...821...88S} and the recently revisited $M_{\rm BH} - PA$ relation by  \cite{Davissubmitted}.  Furthermore, for the measurement of the bulge luminosity, we make use of the \texttt{isofit} task that was recently developed and implemented in the IRAF software package by \cite{2015ApJ...810..120C}.  \texttt{isofit} has dramatically enhanced \citep[e.g.,][]{2016MNRAS.459.1276C} the accuracy of galactic isophotal modeling, and therefore, it substantially increases the accuracy of our calculations. The use of four methods that are largely independent (FP-BH, the $M_{\rm BH} - M_{\rm sph}$, $M_{\rm BH} - n_{\rm sph}$ and the $M_{\rm BH} - PA$ relations) to estimate the BH mass provides an invaluable consistency check. 

The primary goal of this paper, and of our campaign in general, is to discover (or note the absence of) IMBHs in the center of dwarf galaxies with LLAGN. This first part of our campaign has the additional objective of examining the consistency between known $M_{\rm BH}$ in terms of  galactic properties, scaling relations and the FP-BH, in the low mass regime. This is a particularly crucial step, since all scaling relations and the FP-BH have been calibrated primarily in the high-mass regime. For this pilot study, we selected seven LLAGN (based on the detection of strong H$\alpha$ emission lines \citealt{1997ApJS..112..315H}) that have been proposed as hosts of a central IMBH by GS13. For our analysis, we used proprietary radio observations carried out with {\it VLA} between June and September,  2015, archival X-ray and IR observations made with \xmm, \cxo, {\it HST} and {\it Spitzer} telescope, archival FUV and NUV observations by the {\it GALEX} telescope, archival optical observations from the Vatican Advanced Technology Telescope ({\it Vatt}), the Sloan Digital Sky Survey ({\it SDSS}), the Kitt Peak National Observatory ({\it KPNO}) 2.1m telescope, and the 1.0m Jacobus Kapteyn Telescope ({\it JKT}).
In sections \ref{sec-observations} and \ref{sec:results}, we present the details of our data analysis and our results, followed by discussion and our conclusions in sections \ref{discussion} and \ref{conclusion}.

\section{Observations and data analysis}
\label{sec-observations}

\subsection{Radio data}

Between June 21st and September 7th, 2015, we conducted seven radio observations, one for each of our targets, using the Karl G. Jansky Very Large Array ({\it VLA}). Throughout this period, the array was in its most extended A-configuration. We observed in C-band using two 1024-MHz sub-bands, each made up of 512 channels of width 2 MHz, centered at 5.25 GHz and 7.45 GHz. 

Flagging, calibration, and imaging were carried out with the Common Astronomy Software Application \citep[CASA;][]{2007ASPC..376..127M}, using standard procedures (including self-calibration when necessary and possible). Unless otherwise stated in the source-specific comments below, we produced images using \cite{1995AAS...18711202B} weighting with a robust parameter of 0.5 as a trade-off between sensitivity, angular resolution, and side-lobe suppression.

\subsection{X-ray data}
We analyzed all available \cxo and \xmm observations for our source list (see Table~\ref{tab:xray-obs}). Analysis of the \xmm data was done using  
\xmm data analysis software SAS version 14.0.0. and using the calibration files released\footnote{XMM-Newton CCF Release Note: XMM-CCF-REL-331}  on December 1st, 2015. All observations were filtered for high background flaring activity. Following the standard procedure, we extracted high-energy light curves (E$>$10\,keV for MOS and 10$<$E$<$12\,keV for pn) with a 100\,s bin size. Time intervals affected by high particle background were filtered out by placing the appropriate threshold count rates for the high-energy photons.   Spectra were extracted from a circular region centered at the core of each galaxy. In order to achieve the maximum encircled energy fraction\footnote{See \xmm Users Handbook \S3.2.1.1 \\  http://xmm-tools.cosmos.esa.int/external/xmm\_user\_support/ \\ documentation/uhb/onaxisxraypsf.html} within the extraction region and also minimize contamination from extra-nuclear emission, we selected an extraction radius of 30\arcsec. This was reduced to 25\arcsec\ in observations 0112552001 and 0200130101 to avoid including a chip gap and to 20\arcsec\ in observation 0112550101 in order to avoid contamination by adjacent ULX candidate 2XMM J110022.2+285817. The filtering and extraction process followed the standard guidelines 
provided by the \xmm Science Operations Centre (SOC\footnote{http://www.cosmos.esa.int/web/xmm-newton/sas-threads}). Namely, spectral extraction was done with SAS task \texttt{evselect}, using the standard filtering flags (\texttt{\#XMMEA\_EP \&\& PATTERN<=4} for pn and \texttt{\#XMMEA\_EM \&\& PATTERN<=12} for MOS) and SAS tasks \texttt{rmfgen} and \texttt{arfgen} were respectively used to create the redistribution matrix and ancillary file.
In order to achieve the best statistics possible, spectra from all three \xmm CCD detectors (MOS1, MOS2 and pn) were fitted simultaneously for each observation. All detectors, during all observations, were operated in Imaging, Full Frame Mode. With the exception of the pn detector in ObsID 0112552001, where the ``Thick'' filter was used, and all detectors in ObsID 0201690301 where the ``Medium'' filter was used; all other \xmm observations used the ``Thin'' filter.

Spectra from the \cxo  observations were extracted using the standard tools\footnote{http://cxc.harvard.edu/ciao/threads/pointlike/} provided by the latest CIAO software (vers. 4.8.0).
All observations were performed either with the Advanced CCD Imaging Spectrometer I-array (ACIS-I) or the S-array (ACIS-S) in imaging mode.
Starting with the level-2 event files provided by the CXC and CIAO 4.8, we extracted source and background spectra using the CIAO task \texttt{dmextract}. For the \cxo data, spectra were extracted from 3\arcsec  circular area at the center of each galaxy. 
Background spectra were extracted from multiple, source-free regions surrounding the central source. Using the CIAO tasks \texttt{mkacisrmf} and \texttt{mkarf}, we created  the redistribution matrix file (RMF) and the ancillary response file (ARF), respectively. 
To this end, we also employed \texttt{dmstat} to locate the centroids of source and background regions in chip coordinates (information on centroid position is necessary for the RMF and ARF calculation since the calibration varies across the chips) and \texttt{asphist} to create the aspect histogram. The latter is a description of the aspect motion during the observation, which is needed to calculate the ARF.

\subsection{X-ray spectral analysis}

\begin{table*}
 \caption{Best fit parameters for the \cxo and \xmm observations. All errors are in the 90\% confidence range.}
 \begin{center}
\scalebox{0.8}{   \begin{tabular}{lccccccccc}
     \hline\hline\noalign{\smallskip}
     \multicolumn{1}{c}{Source} &
     \multicolumn{1}{c}{ObsID} &
     \multicolumn{1}{c}{Date} &
     \multicolumn{1}{c}{} &
     \multicolumn{1}{c}{} &
     \multicolumn{1}{c}{Exposure time} &
     \multicolumn{1}{c}{nH} &
     \multicolumn{1}{c}{$\Gamma$} &
     \multicolumn{1}{c}{Flux\tablefootmark{a}} &  
     \multicolumn{1}{r}{$\chi^2/dof$ }\\
     \noalign{\smallskip}\hline\noalign{\smallskip}
      \multicolumn{1}{c}{} &
      
     \multicolumn{1}{l}{} &
     \multicolumn{1}{c}{} &     
          \multicolumn{1}{c}{} &

     \multicolumn{1}{c}{} &
               \multicolumn{1}{c}{[ks]} &
     \multicolumn{1}{l}{[$\times10^{21}$\,cm$^2$]} &
     \multicolumn{1}{c}{} &
     \multicolumn{1}{c}{[$10^{-14}\,{\rm erg/cm^2/s}$]} &   
     \multicolumn{1}{c}{} \\
     \noalign{\smallskip}\hline\noalign{\smallskip}

     \noalign{\smallskip}\hline\noalign{\smallskip}

        NGC~628  &   &   &  &   &  &  &    &  &     \\
        &\cxo  2057       & 2001-06-19 &  &  & 46.4\  &  0.46\tablefootmark{b}  & 1.68$_{-0.33}^{+0.38}$ & 2.17$_{-0.36}^{+0.39}$  & 0.61/76    \\        &\cxo  2058       & 2001-10-19 &  &  & 46.2  &  -                    & 1.65$_{-0.33}^{+0.35}$ & 1.66$_{-0.38}^{+1.05}$  & 0.94/83    \\
        &\xmm  0154350101 & 2002-02-01 &  &  & 34.4/36.7/36.7 &            -  & 2.22$_{-0.27}^{+0.25}$ & 2.61$_{-0.50}^{+0.61}$  & 0.95/475   \\
        &\xmm  0154350201 & 2003-01-07 &  &  & 23.1/24.7/24.7          &  -  & 2.03$_{-0.37}^{+0.32}$ & 2.51$_{-0.89}^{+0.32}$  & 0.84/393   \\

        &\cxo  16000      & 2013-09-21 &  &  & 39.5                     &  -  & 1.81$_{-0.54}^{+0.66}$ & 1.48$_{-0.16}^{+0.80}$  & 0.66/38   \\
        &\cxo  16002      & 2013-11-14 &  &  & 37.6                     &  -  & 1.54$_{-0.46}^{+0.49}$ & 1.79$_{-0.72}^{+0.76}$  & 0.80/41   \\
        &\cxo  16003      & 2013-12-15 &  &  & 40.4                     &  -  & 1.52$_{-0.39}^{+0.40}$ & 1.80$_{-0.60}^{+0.62}$  & 0.80/57   \\

      \noalign{\smallskip}\hline\noalign{\smallskip}
     \noalign{\smallskip}\hline\noalign{\smallskip}

        NGC 3185    &    &  &  &    &  &  &  &   &     \\
        &\xmm  0112552001 & 2001-05-07 &  &  & 9.4/14.0/14.0  &  0.21\tablefootmark{b}  & 2.08$_{-0.22}^{+0.24}$ & 4.97$_{-0.88}^{+0.78}$  & 1.01/277   \\
        &\cxo  2760       & 2002-03-14 &  &  &  19.8  &  -                      & 1.86$_{-0.52}^{+0.62}$ & 4.09$_{-1.60}^{+1.59}$  & 0.98/33   \\
      \noalign{\smallskip}\hline\noalign{\smallskip}
      \noalign{\smallskip}\hline\noalign{\smallskip}

        NGC 3198   &    &  &  &    &  &      &  &    &\\
        &\cxo  9551       & 2009-02-07 &  &  &  61.6 &  5.62$_{-3.07}^{+3.58}$  & 1.95$_{-0.53}^{+0.58}$ & 2.75$_{-0.45}^{+0.48}$  & 1.03/88   \\
      \noalign{\smallskip}\hline\noalign{\smallskip}   
     \noalign{\smallskip}\hline\noalign{\smallskip}

        NGC 3486    &    &  &  &    &  &     &  &    & \\
        &\xmm  0112550101 & 2001-05-09&  &  & 10.0/9.0/8.9  &  0.19\tablefootmark{b}  & 2.06$_{-0.32}^{+0.34}$  & 5.64$_{-2.24}^{+2.26}$  & 0.76/262   \\
      \noalign{\smallskip}\hline\noalign{\smallskip}  
      \noalign{\smallskip}\hline\noalign{\smallskip}

        NGC 3507\tablefootmark{c}   &    &  &  &    & &  &    &  &     \\        &\cxo  3149      & 2002-03-08 &  &  &  39.3   &  0.15\tablefootmark{b} & 2.06$_{-0.85}^{+0.61}$ & 1.11$_{-0.51}^{+0.46}$  & 1.25/79   \\
      \noalign{\smallskip}\hline\noalign{\smallskip}   
      \noalign{\smallskip}\hline\noalign{\smallskip}

        NGC 4314\tablefootmark{d}   &       &  &  & &  &  &    &  &     \\
        &\cxo  2062       & 2001-04-10 &  &  &  16.1  &  0.57\tablefootmark{b} & 1.54$_{-1.38}^{+2.01}$ & 1.86$_{-1.42}^{+1.57}$  & 1.01/64   \\

        &\cxo  2063       & 2001-07-31&   &   & 16.0  &  -                     & 2.12$_{-0.89}^{+0.71}$ & 1.92$_{-1.07}^{+0.98}$  & 0.92/56   \\

        &\xmm  0201690301 & 2004-06-20 &  &  & 30.0/31.7/31.7 &  -             & 1.87$_{-0.27}^{+0.28}$ & 7.05$_{-1.06}^{+1.09}$  & 0.89/237   \\
      \noalign{\smallskip}\hline\noalign{\smallskip}         
      \noalign{\smallskip}\hline\noalign{\smallskip}

        NGC 4470    &    &  &  &  &  &    &  &       & \\
        &\cxo  321        & 2000-06-12&  &   &   39.6          &  0.17\tablefootmark{b} & 1.75$_{-0.31}^{+0.34}$ & 3.89$_{-1.15}^{+1.16}$  & 1.05/88   \\
        &\xmm  0200130101 & 2004-01-02&  &   & 91.9/94.5/94.5  &  -                     & 2.34$_{-0.25}^{+0.26}$ & 3.39$_{-0.51}^{+0.23}$  & 0.93/578   \\
        &\cxo  12978      & 2010-11-20&  &   &  19.8                               &  - & 1.99$_{-0.49}^{+0.58}$ & 3.63$_{-1.21}^{+1.23}$  & 0.90/57   \\
        &\cxo  12888      & 2011-02-21&  &   &  159.3                              &  - & 1.92$_{-0.44}^{+0.48}$ & 3.77$_{-1.18}^{+1.19}$  & 1.02/67   \\
        &\cxo  15756      & 2014-04-18&  &   &  32.1                               &  - & 2.09$_{-0.64}^{+0.84}$ & 3.32$_{-1.20}^{+1.21}$  & 0.91/35   \\
        &\cxo  15760      & 2014-04-26&  &   &  29.4                               &  - & 2.28$_{-0.69}^{+0.76}$ & 3.35$_{-1.27}^{+1.29}$  & 1.06/24   \\
        &\cxo  16260      & 2014-08-04&  &   &  24.7                               &  - & 2.36$_{-0.61}^{+0.55}$ & 2.11$_{-0.62}^{+0.73}$  & 0.78/39   \\
        &\cxo  16261      & 2015-02-24&  &   &  22.8                               &  - & 2.15$_{-0.49}^{+0.54}$ & 2.17$_{-0.72}^{+0.74}$  & 0.82/51   \\

      \noalign{\smallskip}\hline\noalign{\smallskip}         
          
      \noalign{\smallskip}\hline\noalign{\smallskip}
    \end{tabular}   }
 \end{center}
  \tablefoot{
  \tablefoottext{a}{Unabsorbed, 0.5-10\,keV.}\\
  \tablefoottext{b}{Parameter frozen at total galactic H\,I column density \citep{1990ARA&A..28..215D}.} \\
  \tablefoottext{c}{With an additional emission component from hot diffuse gas, modeled using the {\small {XSPEC}} model \texttt{mekal} for a plasma temperature of 0.57$_{-0.07}^{+0.06}$\,keV.} \\
  \tablefoottext{d}{With an additional emission component from hot diffuse gas, modeled using the {\small {XSPEC}} model \texttt{mekal} for a plasma temperature of 0.33$_{-0.02}^{+0.03}$\,keV for the \cxo 2062 and \xmm observations and 0.45$_{-0.12}^{+0.13}$\,keV for \cxo 2063.}}
 \label{tab:xray-obs}
\end{table*}

All spectra were regrouped to have at least one count per bin and analysis was performed using the {\tt xspec} spectral fitting package, version 12.9.0 \citep{1996ASPC..101...17A}, employing Cash statistics \citep{1979ApJ...228..939C}. 
An absorbed power law model was used in
the analysis of all observations. Further, spectra were fitted with the {\tt xspec} model {\tt tbnew*po}, where {\tt tbnew} is the latest improved version of the  {\tt tbabs} X-ray absorption model
\citep{2011_in_prep}. In two cases, the central core spectra were contaminated by galactic diffuse emission, which was fitted using a single temperature \texttt{mekal} model \citep{1985A&AS...62..197M,1986A&AS...65..511M,Kaastra_1992,1995ApJ...438L.115L}. We only considered galactic interstellar absorption in the direction of each source \citep{1990ARA&A..28..215D}, with the exception of NGC 3198, for which we obtained strong indications of intrinsic absorption. The results of our analysis are tabulated in Table~\ref{tab:xray-obs} and specifics of each source are briefly presented in section~\ref{specific}.

\subsection{ Source specific comments -- X-ray \& radio}
\label{specific}

 {\it NGC~628} is located at a distance (distances for all sources are presented in Table~\ref{sec-spec_x_III}) of $\sim10.2$\,Mpc \citep{2014ApJ...792...52J}. 
 The source has been observed multiple times by both \cxo and \xmm between 2001 and 2013. Due to its proximity and small inclination \citep{1984A&A...132...20S}, NGC 628 has proven to be an ideal backdrop for the discovery of multiple extragalactic X-ray sources, in addition to its central engine \citep[e.g.,][]{2002ApJ...572L..33S}. For this work, we analyzed the spectra of the five longest \cxo observations and the two available \xmm observations. Spectra from all observations were fitted with an absorbed power law model. The best fit value for the photon index ranged between 1.52$_{-0.39}^{+0.40}$ and 2.22$_{-0.27}^{+0.25}$, throughout the different observations. Obviously, there is a noticeable variation in the central value of the photon index. Nevertheless, this is due to the low number of total registered photons, rather than an actual modification of the state of the source. Namely, the small number of counts yields moderately high uncertainties for all best fit values. Indeed, despite this apparent variation, all fit values are consistent within the 90\% confidence range. The source flux does not appear to vary significantly (again all values are consistent within 90\% errors) between the different observations. The mean unabsorbed luminosity of the central source in the 0.5-10\,keV range is $\approx2.49\times10^{38}$\,erg s$^{-1}$. 
 
 No radio source was detected at the position of NGC~628 in any of the sub-bands using \cite{1995AAS...18711202B} weighting down to a root mean square (RMS) noise level of $8.4 \, \mu$Jy. To improve sensitivity, we combined both bands to produce a naturally weighted image, which did not reveal any radio source but provided a more constraining RMS noise level of $5.5 \, \mu$Jy and a corresponding 3$\sigma$ upper limit of $16.5 \, \mu$Jy. The corresponding radio luminosity ($L_R$), along with all values or upper limits for the luminosities of all sources in our sample, are presented in column 6 of Table~\ref{sec-spec_x_III}.

{\it NGC 3185} is  located  at 25.8\,Mpc \citep{2007A&A...465...71T}. There are two observations of the source in the X-ray band. Spectra extracted from both observations were fitted with an absorbed power law model, with a photon index ranging between 1.9 and 2.1 (best fit value for \cxo and \xmm respectively). The mean unabsorbed source luminosity is $\approx3.61\times10^{39}$\,erg s$^{-1}$. 

We detect a faint radio source, coincident with the optical nucleus of  NGC 3185, in both sub-bands. The source is unresolved to a beam size of $0.48 \times 0.24$ arcsec at 7.45 GHz. We measure the following flux densities by fitting a gaussian model to the source in the image plane: $64.7 \pm 6.1 \, \mu$Jy and $46.0 \pm 5.4 \, \mu$Jy at 5.25 GHz and 7.45 GHz, respectively.

 {\it NGC 3198} is located at a distance of 13.8\,Mpc \citep{2001ApJ...553...47F}. Its spectrum is modeled with an absorbed power law, with the column density as a free parameter. The fit reveals a column density that is approximately 55 times higher than what is expected from Galactic absorption and can only be attributed to intrinsic obscuration. This finding is in agreement with previous classification of NGC 3198 as an obscured AGN \citep{2011ApJ...731...60G}. The 0.5-10\,keV source unabsorbed luminosity is $\approx6.27\times10^{38}$\,erg s$^{-1}$.

 A radio point source (beam size of $0.28 \times 0.26$ at 7.45 GHz) is detected in both frequency bands. The position of the radio source is consistent with the optical nucleus of NGC 3198. We measure the following flux densities: $118 \pm 12 \, \mu$Jy and $116 \pm 9 \, \mu$Jy at 5.25 GHz and 7.45 GHz, respectively.

{\it NGC 3486} is found at a distance of 16.7\,Mpc \citep{2007A&A...465...71T}. Again the source spectrum was fitted with an absorbed power law. The source unabsorbed luminosity was $\approx1.01\times10^{39}$\,erg s$^{-1}$ in the 0.5-10\,keV range.

A marginal 4$\sigma$ detection is obtained for the radio counterpart of NGC 3486 by combining the two frequency bands and using natural weighting. The source is point-like and coincident with the optical core of the galaxy. We measure a flux density of $26.6 \pm 6.4 \, \mu$Jy.

\begin{figure}
       \resizebox{\hsize}{!}{\includegraphics[angle=-90,clip,trim=0 0 0 0]{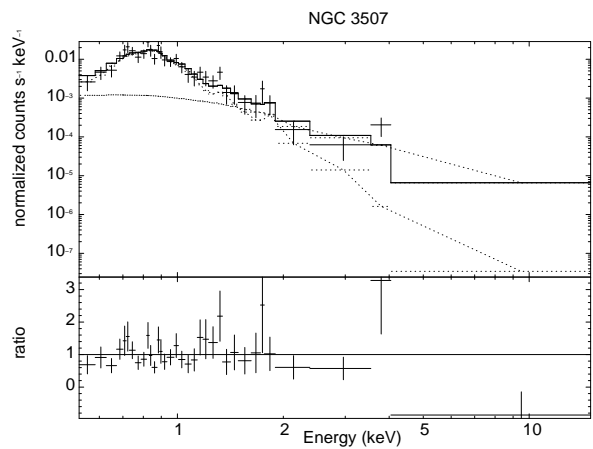}}
            \caption{NGC 3507: registered counts per keV vs. energy and data-vs-model ratio plot, for the {\tt tbnew*(mekal+po)} model. The continuum emission is clearly not described by a power law spectrum, resulting in strong residuals above 3\,keV. }
   \label{fig:res}
 \end{figure}

{\it NGC 3507} is located at a distance of $\sim$16.3\,Mpc. We extracted the spectrum from the central core and  modeled it using an absorbed power law. The resulting fit is unacceptable with a reduced $\chi^2>3$. Adding an emission component originating in hot diffuse gas using the {\tt xspec} model \texttt{mekal} with a plasma temperature of $\sim0.57$\,keV improves the fit, yielding a  reduced $\chi^2$ of 1.25 for 79 dof. However, the fit is still unacceptable and there are strong residuals above 3\,keV (see Figure~\ref{fig:res}). It appears that the X-ray continuum is not well described by a power law. 
This issue is examined further in the discussion section. The 0.5-10\,keV unabsorbed luminosity of the central region (based on the power law component for the above fit) is $\approx3.53\times10^{38}$\,erg s$^{-1}$ (total source luminosity including a diffuse, extra-nuclear component $\approx1.07\times10^{39}$\,erg s$^{-1}$).

We detect a radio counterpart to the optical nucleus of NGC 3507 in both sub-bands. Gaussian fitting in the image plane indicates that the source is marginally resolved. We indeed notice a slight extension of the emission to the south that can be distinguished more easily at 7.45 GHz due to the better angular resolution.  We measure the following flux densities: $203 \pm 13 \, \mu$Jy and $132 \pm 14 \, \mu$Jy at 5.25 GHz and 7.45 GHz, respectively. However, since the structure of the source suggests a possible contamination by emission components other than the jets, our measured jet flux densities are likely overestimated.

{\it NGC 4314} is located at a distance of 13.1\,Mpc \citep{2002AJ....123.1411B}. The spectra of all three available observations were fitted by an absorbed power law with a photon index of approximately two plus a diffuse X-ray emission component from ionized gas, modeled using a single temperature \texttt{mekal} model, with T$\sim0.4$\,keV. 
The  0.5-10\,keV mean value for the source unabsorbed luminosity -- calculated for the 3\arcsec\ extraction region from the \cxo observations, and only for the power law component -- is $\approx3.88\times10^{38}$\,erg s$^{-1}$ ; total 0.5-10\,keV source luminosity for the entire 30\arcsec\ region (for all \cxo and \xmm observations), is $\approx1.38\times10^{39}$\,erg s$^{-1}$. 

As already mentioned by \cite{1991A&A...244..257G}, a ring of radio emission is (marginally) detected around the position of the galaxy nucleus. However, we do not detect radio emission at a position coincident with the optical core \citep[also in agreement with][]{1991A&A...244..257G}. Combining our two spectral bands and using natural weighting, we obtain a RMS noise level of $7.6 \, \mu$Jy leading to a 3$\sigma$ upper limit of  $23 \, \mu$Jy.

 {\it NGC 4470} has been observed multiple times by both \cxo and \xmm. Here we analyze  one \xmm and two \cxo observations that provided a sufficient number of counts for a reliable fit. The observation dates range from 2000 to 2015. In all observations the spectrum was fitted with an absorbed power law. The best fit value of the power law photon index ranged between 1.75$_{-0.31}^{+0.34}$ to 2.36$_{-0.61}^{+0.55}$ throughout the observations, with all values being consistent, in the 90\% range. The source unabsorbed luminosity (0.5-10\,keV) varied moderately between different observations, ranging between 3.04$_{-0.89}^{+1.05}$\,erg s$^{-1}$ in 2014 and 5.57$_{-1.69}^{+1.70}$\,erg s$^{-1}$ in 2000, for a distance of $\sim$34.7\,Mpc \citep{2007A&A...465...71T}. Again, the different luminosity estimates are consistent within the 90\% confidence range.
 
 No radio source was detected at the position of NGC 4470 on a naturally weighted image using the full spectral range. We obtain a 3$\sigma$ upper limit of $21 \, \mu$Jy.

\subsection{Optical \& IR data}
\label{opt-IR}

Due to the non-trivial morphologies of these late-type galaxies, with many displaying features such as bars, rings and spiral arms, we preferred high-angular-resolution imaging so as to ensure that all the structural components (particularly the small-scale spheroids hosting the central BHs) are well resolved and readily separable from the other galactic components in the radial surface brightness profiles. We therefore opted for archival {\it Hubble Space Telescope (HST)} data\footnote{http://hla.stsci.edu} taken with the $WFPC2$ or $ACS/WFC$ cameras, for the galaxies NGC~628, NGC~3185, and NGC~3486. We used images taken with the near infrared (NIR) filter $F814W$ in order to minimise obscuration from dust. 

The remaining galaxies in our sample, however, either displayed significant dust in this band (often crossing the bulge photometric center), or their radial extent exceded the $HST$ detector's field-of-view (and consequently had over-subtracted sky background via the data reduction pipeline that provided these images), or both.  Therefore, for these galaxies (NGC~3198, NGC~3507, NGC~4314 and NGC~4470) we used longer wavelength and wider field-of-view imaging retrieved from the {\it Spitzer Survey of Stellar Structure in Galaxies (S$^4$G)} archive\footnote{http://irsa.ipac.caltech.edu/data/SPITZER/S4G/}, taken with the $IRAC~1$ instrument, at 3.6\,$\mu$m. 

The optical data used for measuring pitch angles were obtained from the NASA/IPAC Extragalactic Database (NED). We used the R-band imaging data taken from various sources. When given the option, the imaging with the best resolution was used, specifically NGC~628 from the {\it Vatt}, NGC~3507 from the Palomar 48-inch Schmidt, NGC~4470 from the {\it SDSS}, NGC~3198 from the {\it KPNO}, NGC 3486 from the {\it Bok Telescope}, and NGC~3185 and NGC~4314 from the {\it JKT}.

\subsection{IR imaging Analysis}

For each galaxy image, we masked out the contaminating sources, such as globular clusters, foreground stars, background galaxies and, where applicable, dust. We further characterized the instrumental point spread function (PSF) of each image by fitting Moffat profiles to several bright, unsaturated stars, with the IRAF task {\tt imexamine}. 

The diffuse galaxy light was modeled by fitting quasi-elliptical isophotes as a function of increasing semi-major axis, with the $IRAF$ task {\tt isofit}. This yielded the major axis surface brightness profile, $\mu(R_{\rm maj})$, which was decomposed with the software {\tt profiler} \citep{2016arXiv160708620C}. In short, {\tt profiler} constructs a model radial profile using pre-defined analytical functions (S\'ersic, Gaussian, exponential, etc.) that describe particular photometric components (e.g., disc, bulge, bar, point-source).  All the chosen components are added together and the result is convolved with the PSF, and this is iterated by varying the parameters of each component until the best-fitting solution is found.

We further mapped the major axis profiles onto the geometric mean (or `equivalent') axis, $R_{\rm eq}$, defined as $R_{\rm eq} = R_{\rm maj} \sqrt{1 - \epsilon(R_{\rm maj})}$, with $\epsilon(R_{\rm maj})$ being the isophote ellipticity at each radius. This mapping `circularizes' each isophote such that the enclosed surface area is conserved (see the Appendix in \citealt{2015ApJ...810..120C} for further details). Decomposing $\mu(R_{\rm eq})$ allows for an analytical calculation of each component's total magnitude. 

An example decomposition is shown in Figure~2, for NGC~628. For each galaxy, the spheroid major axis S\'ersic index ($n_{\rm sph}$) and absolute magnitude at 3.6\,$\mu$m ($M_{{\rm sph}}$) resulting from the analysis, are listed in Table \ref{sec-spec_x_III}. Full decompositions for all seven galaxies are presented in Ciambur et al. (in prep.), which provides an in-depth analysis of the IMBH candidate galaxies from GS13 that host AGN but have low bulge luminosities.

\begin{figure*}
\includegraphics[width=.51\textwidth]{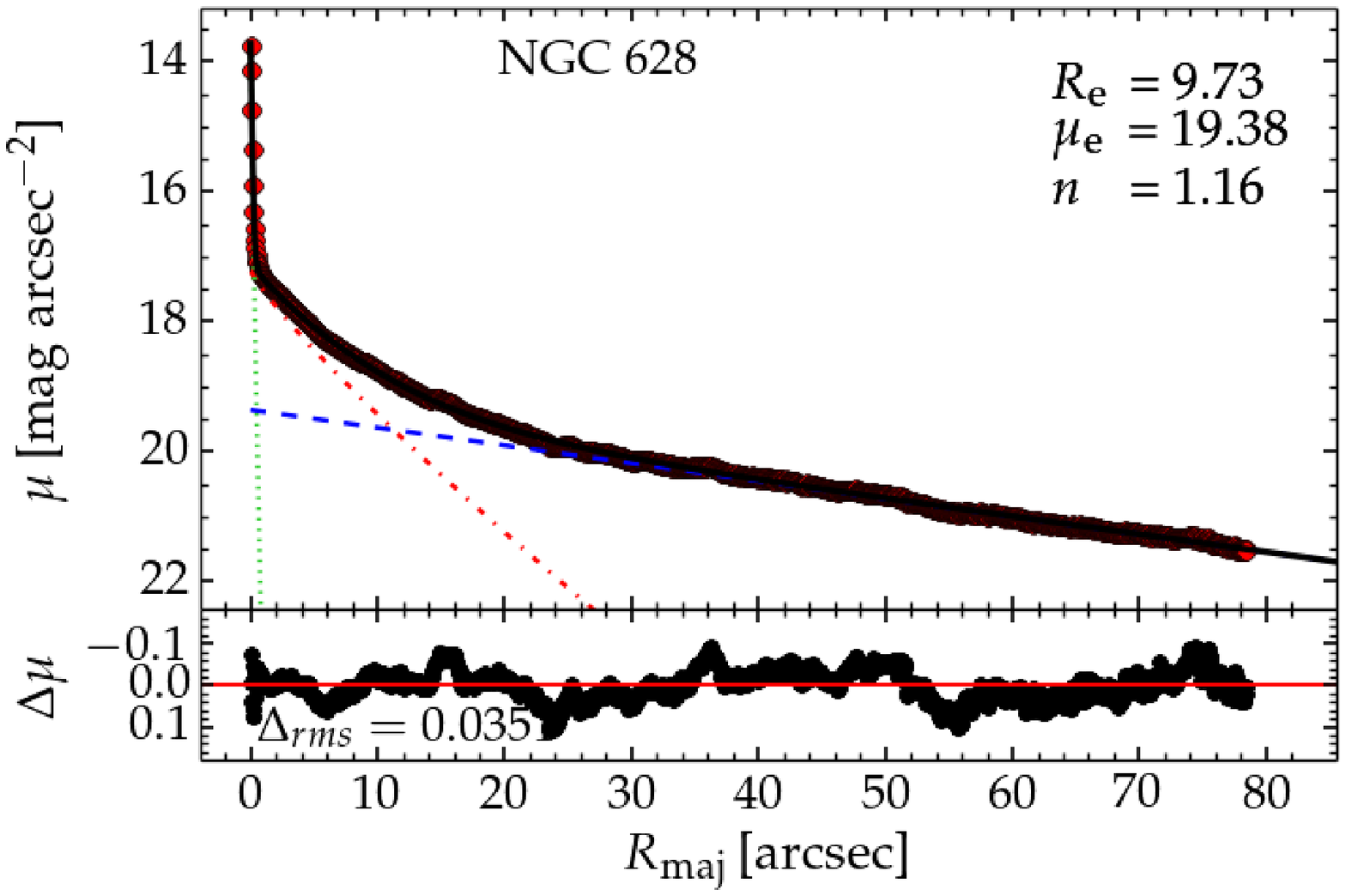}
\includegraphics[width=.51\textwidth]{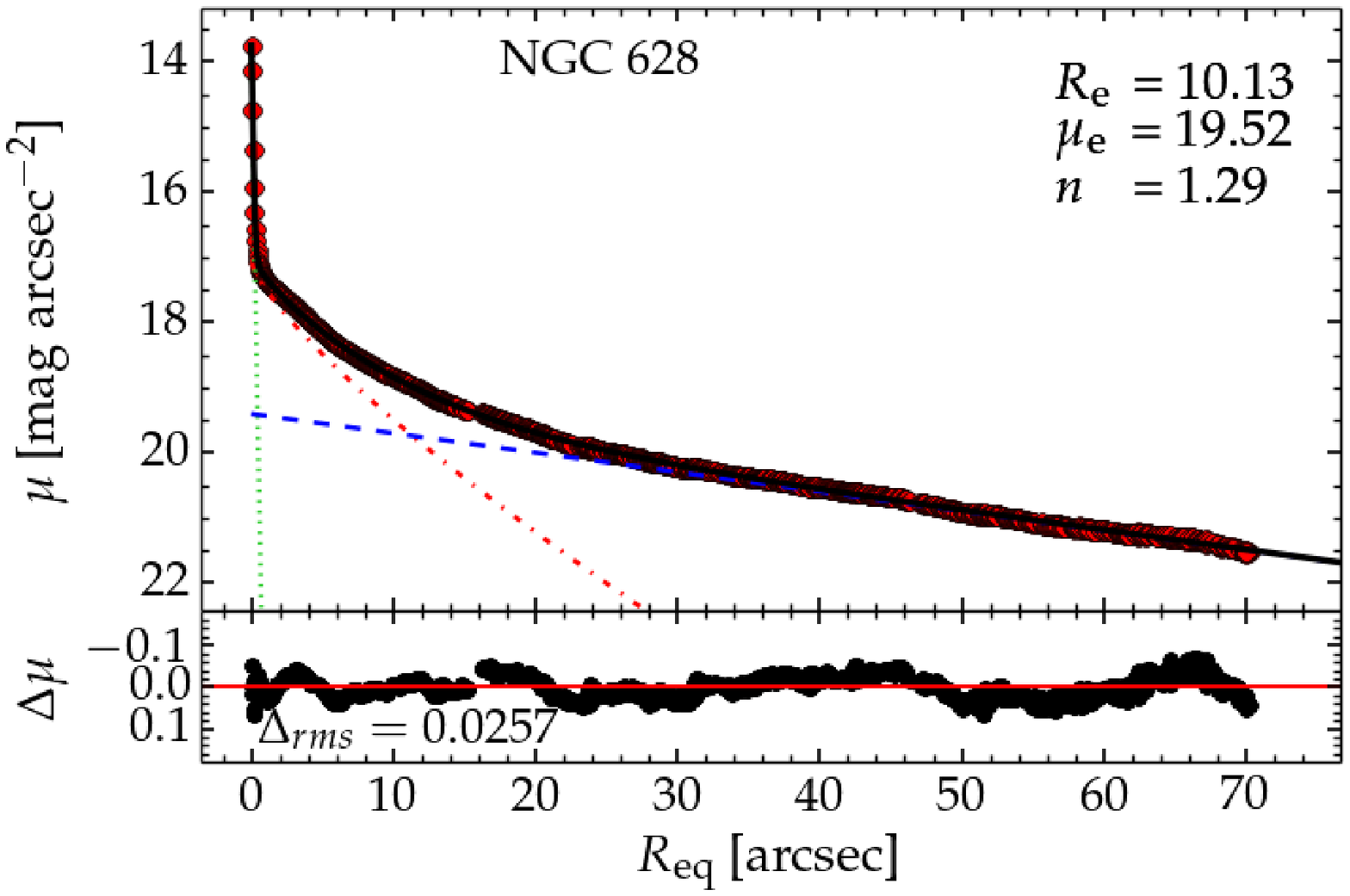}
\label{fig:decomp}
\caption{Top panels show the decomposed surface brightness profile of NGC~628 along the major (left) and equivalent (right) axis. The data is represented as red circles while the model is the black curve. This galaxy has a central point source (green, dotted line), a S\'ersic spheroid (red, dash-dotted line) and an exponential disc (blue, dashed line). Inset in the upper left are the best-fit spheroid parameters. Below the decomposition panels are the residual ($\Delta\mu$) profiles.}

\end{figure*}

\subsection{Source specific comments -- IR}

$NGC~4470 : $ we find that NGC~4470 is a bulgeless spiral galaxy. Therefore, we can not estimate the $M_{\rm BH}$ from the BH - bulge scaling relations. There is a point source very close ($\sim 0.5$ arcsec) to the disc's photometric centre; possibly the emission coming from the LLAGN in the IR. The profile is well modeled by only an exponential disc and a bar component (the latter probably being previously mistaken as the bulge by D\&D06, who report a S\'ersic index of 0.7, that is, consistent with a bar profile).\\

\subsection{Spiral arms analysis}

We use the two-dimensional Fast Fourier Transform (\texttt{2DFFT}), which is a widely used method to obtain pitch angles of spiral arms \citep{1994A&AS..108...41S,1996ApJ...460..651G,2008ApJ...678L..93S,2012ApJS..199...33D, 2014ApJ...793L..19M}. It allows for decomposition of spiral
arms into sums of logarithmic spirals of varying pitches. The method described in \cite{1994A&AS..108...41S} measures both the strength and the $PA$ of the various modes between a given inner radius and a given outer radius on a deprojected image. The extension to this method in \cite{2012ApJS..199...33D} eliminates the user defined inner radius and allows the user to quantitatively investigate  how logarithmic the spiral arms are. This provides a systematic way of excluding barred nuclei from the pitch angle measurement annulus. We therefore opted for the method of \cite{2012ApJS..199...33D}, which outputs the $PA$ as a function of inner radius.

The images were first deprojected to face-on by assuming that the disk galaxy, when face-on, will have circular isophotes. The galaxy axis ratios were determined using SExtractor \citep[Source Extractor;][]{1996A&AS..117..393B}. Then, we seek a harmonic mode in which we find the largest range of inner radii over which the $PA$ stays approximately constant. The $PA$ is measured by averaging over this stable region, with the error being determined by considering the size of the stable radius segment relative to the galaxy radius, as well as the degree to which the stable segment is logarithmic. Errors in the determination of the galaxy center or galaxy axial ratio in the deprojection process are not critical to the measurement of the $PA$ \citep[see][for more detail]{2012ApJS..199...33D}.

\begin{figure*}
\includegraphics[width=.51\textwidth]{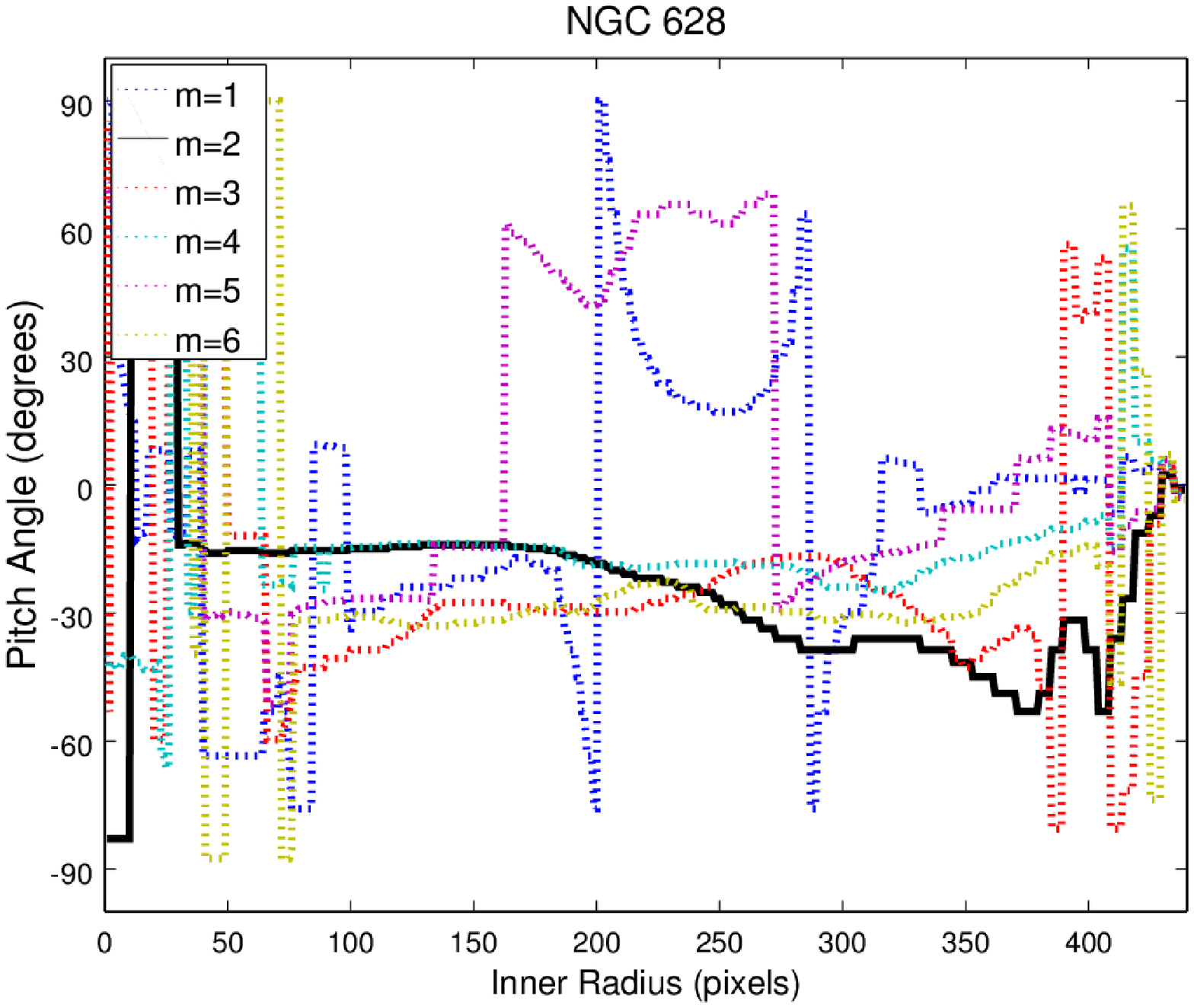}
\hspace{1.0cm}
\includegraphics[width=.38\textwidth, scale=0.35]{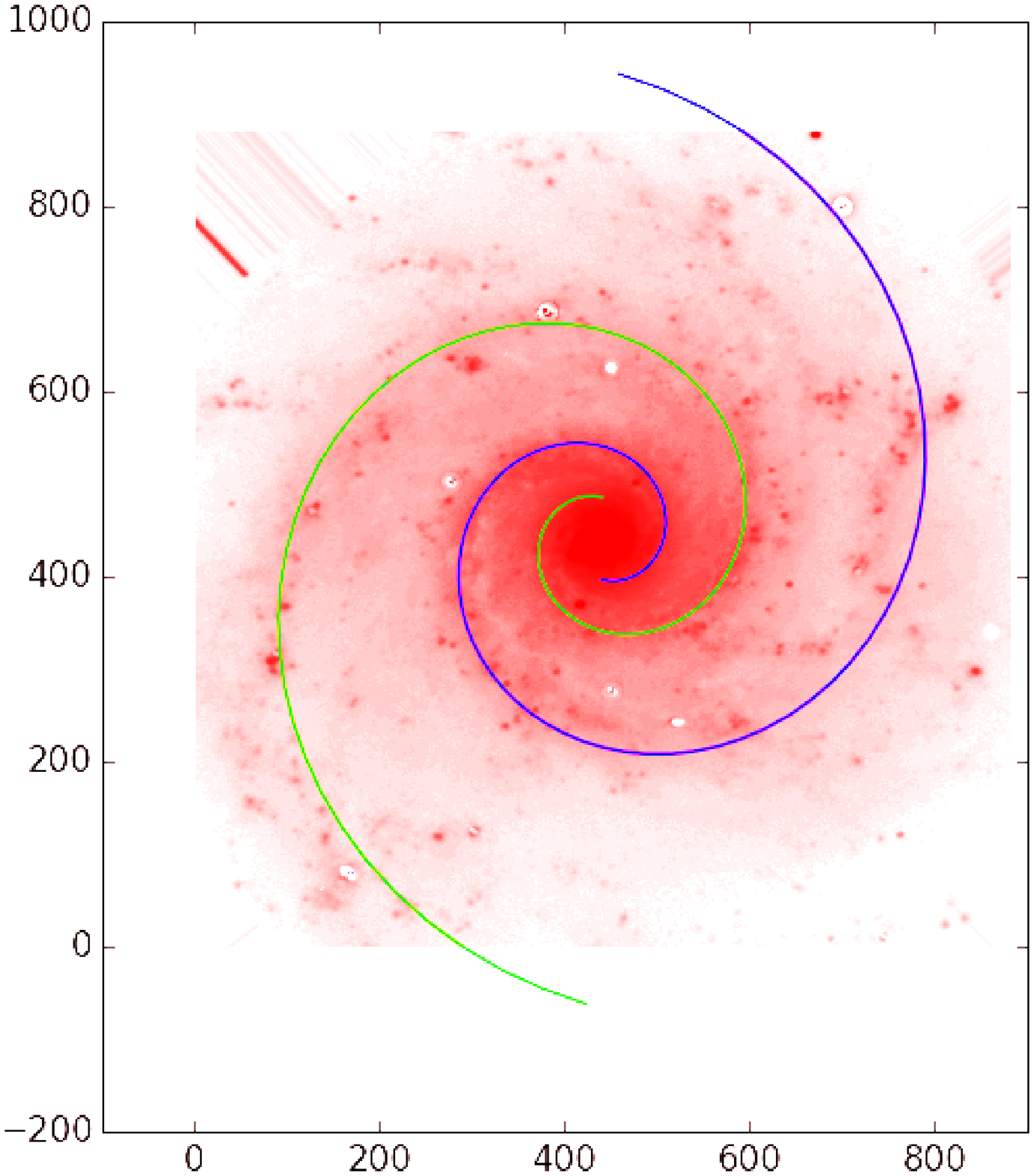}
\caption{ {\it Left}: NGC~628 R-band $PA$ as a function of inner radius for deprojected image. A stable mean $PA$ is determined for the m=2 harmonic mode (black solid line) from a minimum inner radius of 30 pixels (12\arcsec) to a maximum inner radius of 200 pixels (80\arcsec), with an outer radius of 440 pixels (176\arcsec). The pitch angle can be seen to decrease dramatically with increasing radius beyond approximately 50\% of the outermost radius and becoming chaotic near the edge; this is due to a shrinking annulus of measurement reducing the viability of pitch angle measurement in the outer regions of the galaxy. {\it Right}:  an overlay image of a 14.98\,degrees pitch angle spiral on top of NGC~628.}
\label{arms}

\end{figure*}

An example deprojection and measurement of the spiral $PA$ for NGC~628 is shown in Figure~\ref{arms}. The source was measured to have a position angle of -40.5 degrees and elongation of 1.103 using \texttt{SExtraxtor}. The image was deprojected after a Gaussian star subtraction was performed to reduce the noise from the foreground stars. Although star subtraction is a helpful step, the presence of foreground stars does not affect the value of the $PA$ in general \citep[][]{2012ApJS..199...33D}. The plot shows the output of the \texttt{2DFFT} code, which is the $PA$ as a function of inner radius for the deprojected image.  A stable mean $PA$ is determined for the m=2 harmonic mode from a minimum inner radius of 30 pixels (12{\arcsec}) to a maximum inner radius of 200 pixels (80{\arcsec}), with an outer radius of 440 pixels (176{\arcsec}). Equation 7 in \cite[][]{2012ApJS..199...33D} describes the final error in $PA$ measurement: $E_{\phi}=(( \beta * \sigma / \lambda )^2+( \epsilon_{m} )^2)^{(1/2)}$, where $E_{\phi}$ is the total $PA$ error, $\epsilon_{m}$ is the quantized error for the dominant harmonic mode, $\sigma$ is the standard deviation about the mean $PA$, $\beta$ is the distance in pixels from the innermost stable spiral structure to 90\% of the selected outer radius of the galaxy ($0.9*r_{max}$), and $\lambda$ is the length of the stable range of radii over which the $PA$ is averaged. For NGC~628, the equation yields $E_{\phi}=2.23$ degrees with $\lambda=170 pixels$, $\beta=366 pixels$, $\sigma= 0.95$ degrees, and $\epsilon_{2}$=0.88 degrees. The final determination of $PA$ is therefore $-15.07 \pm 2.23$.  We also checked our measurement with the \texttt{Spirality} code of \cite{2015arXiv151106365S}, which uses template fitting rather than Fourier decomposition to measure $PA$. The results from these two independent methods agree well within the uncertainties. Finally, we generated an overlay image to double check our measurement by visualizing the spiral on top of the galaxy and confirming the result by eye (see Fig~\ref{arms}). The same procedure was used for all seven galaxies in our sample. The $PA$ estimations are presented in column 5 of Table~\ref{sec-spec_x_III}

\section{Results}
\label{sec:results}

 \begin{sidewaystable*}
 \caption{Mass of the central black hole, using $M_{\rm BH}$-spheroid relations by \cite{2013ApJ...764..151G}, \cite{2016ApJ...821...88S} and \cite{2016ApJ...817...21S}, the $M_{\rm BH}-PA$ relation by  \cite{Davissubmitted} and the FP-BH by \cite{2012MNRAS.419..267P}. All errors are 1$\sigma$. }
 \begin{center}
\scalebox{0.8}{   \begin{tabular}{lccccccccccccc}
     \hline\hline\noalign{\smallskip}
     \multicolumn{1}{c}{Source} &
     \multicolumn{1}{c}{Morphology} &
     \multicolumn{1}{c}{Dist.\tablefootmark{a}} &
     \multicolumn{1}{c}{$n_{\rm sph}$ } &
     \multicolumn{1}{c}{ $M_{{\rm sph}}$} &
     \multicolumn{1}{c}{ $PA$} &
     \multicolumn{1}{c}{$L_R$\tablefootmark{b}} &
          \multicolumn{1}{c}{$L_X$\tablefootmark{c}} &
     \multicolumn{1}{c}{log($M_{\rm BH}$)\tablefootmark{d}} &
     \multicolumn{1}{c}{log($M_{\rm BH}$)\tablefootmark{e}} &
     \multicolumn{1}{c}{${\rm log}(M_{\rm BH})$\tablefootmark{f}} &
     \multicolumn{1}{c}{${\rm log}(M_{\rm BH})$\tablefootmark{g}} &
     \multicolumn{1}{c}{${\rm log}(M_{\rm BH})$\tablefootmark{h}} &
     \multicolumn{1}{c}{} \\
     
     \noalign{\smallskip}\hline\noalign{\smallskip}
      \multicolumn{1}{l}{} &
     \multicolumn{1}{c}{} &
     \multicolumn{1}{c}{} &
       \multicolumn{1}{c}{} &
              \multicolumn{1}{c}{} &
      \multicolumn{1}{c}{} &
     \multicolumn{1}{c}{} &
          \multicolumn{1}{c}{} &
          \multicolumn{1}{l}{} &
     \multicolumn{1}{c}{}&
          \multicolumn{1}{c}{}&
     \multicolumn{1}{c}{}&
     \multicolumn{1}{c}{}  \\
     
     \noalign{\smallskip}\hline\noalign{\smallskip}
      \multicolumn{1}{c}{} &
            \multicolumn{1}{c}{} &
     \multicolumn{1}{l}{[Mpc]} &
          \multicolumn{1}{c}{} &
      \multicolumn{1}{c}{[@$3.6\,\mu$]} &
            \multicolumn{1}{c}{} &
     \multicolumn{1}{c}{[$10^{34}\,{\rm erg/s}$]} &
          \multicolumn{1}{c}{[$10^{38}\,{\rm erg/s}$]}&
     \multicolumn{1}{l}{[FP-BH]} &
     \multicolumn{1}{c}{[GS13] } &
     \multicolumn{1}{c}{[$M_{\rm BH} - n_{\rm sph}$ ] } &
     \multicolumn{1}{c}{[$M_{\rm BH} - M_{\rm sph}$] } &
          \multicolumn{1}{c}{[$M_{\rm BH} - PA$] }  \\
         
     \noalign{\smallskip}\hline\noalign{\smallskip}

     \noalign{\smallskip}\hline\noalign{\smallskip}

NGC~628  & SA(s)c  & 10.2$\pm1.05$& 1.16$\pm$0.20 & -20.57$\pm$0.33  &  -15.07$\pm$2.23 &$<$1.03          & $2.49 \rm{ \pm0.90}$       &  $<7.5$               & 4.9$\pm1.0$     & 6.7$\pm$0.4  & 6.6$\pm$0.7   &$7.0 \pm 0.4$ \\
                                                                                                                                                                                                        
NGC~3185 & SBc     & 25.8$\pm5.33$& 1.77$\pm$0.25 & -20.27$\pm$0.41  &  -16.54$\pm$3.33 & 25.9$\pm$10.9   & $36.1 \rm{ \pm16.9}$       & 6.5$\pm2.1$         & 5.3$\pm$0.9      & 7.3$\pm$0.3   & 6.6$\pm$0.8   &$6.9 \pm 0.5$    \\
                                                                                                                                                                                                        
NGC~3198 & SB(rs)c & 13.8$\pm0.38$& 1.08$\pm$0.35 & -20.06$\pm$0.32  &   27.17$\pm$6.10 & 13.4$\pm1.56$   & $6.27 \rm{ \pm1.08}$       & 6.9$\pm2.1$         & 4.4$\pm1.0$      & 6.6$\pm$0.5    & 6.0$\pm$0.8  &$5.5 \pm 0.8$    \\
                                                                                                                                                                                                        
NGC~3486 & SAB(r)c & 16.7$\pm3.07$& 2.43$\pm$0.40 & -21.44$\pm$0.50  &   12.98$\pm$0.69 & 4.44$\pm1.95$   & $18.8 \rm{ \pm10.1}$       & 5.6$\pm2.1$          & 4.3$\pm1.0$      & 7.7$\pm$0.3    & 7.0$\pm$0.6 &$7.3 \pm 0.9$    \\
                                                                                                                                                                                                          
NGC~3507 & SB(s)b  & 16.3$\pm1.10$ & 1.74$\pm$0.35& -20.29$\pm$0.53  &  -14.51$\pm$1.26  & 22.3$\pm0.36$   & $3.53 \rm{ \pm1.66}$       & 7.5$\pm2.1$          & 5.4$\pm0.9$      & 7.3$\pm$0.3    & 6.1$\pm$0.8&$7.1 \pm 0.3$    \\
                                                                                                                                                                                                        
NGC~4314 & SB(rs)a & 13.1$\pm0.91$& 1.20$\pm$0.28 & -20.89$\pm$0.43  &   28.59$\pm$6.74 & $<$2.66         & $3.88 \rm{ \pm 2.68}$      &$<7.6$                & 5.5$\pm$0.9      & 6.8$\pm$0.5    & 6.4$\pm$0.7 &$5.3 \pm 0.9$    \\
                                                                                                                                                                                                        
NGC~4470 & SAc     & 34.7$\pm6.39$& --            & --               &   11.62$\pm$3.70 & $<$15.6         & $46.1 \rm{ \pm 18.2}$      & $<7.9$                 & 4.9$\pm$1.0     &--              & --         &$7.5 \pm 0.5$    \\

      \noalign{\smallskip}\hline\noalign{\smallskip}         
          
      \noalign{\smallskip}\hline\noalign{\smallskip}
    \end{tabular}   }
 \end{center}
  \tablefoot{$M_{\rm BH}$ estimations along with distance, S\'ersic index ($n_{\rm sph}$), spheroid magnitude ($M_{\rm sph}$), pitch angle ($PA$), radio ($L_R$) and X-ray ($L_X$) luminosity.\\
      \tablefoottext{a}{References for distance estimations for each source are given in Section~\ref{specific}} \\
      \tablefoottext{b}{Calculated at the frequency of 5.25\,GHz.} \\
      \tablefoottext{c}{In the 0.5-10\,keV range. For sources with multiple observations, we use the mean of all values provided by the best fits (i.e. Table~\ref{tab:xray-obs}) } \\
      \tablefoottext{d}{In units of $M_{\odot}$, using the FP-BH by \cite{2012MNRAS.419..267P}}\\
      \tablefoottext{e}{In units of $M_{\odot}$, using the $M_{\rm BH}-L_{\rm sph}$ relation by   \cite{2013ApJ...764..151G}}\\
      \tablefoottext{f}{In units of $M_{\odot}$, using the $M_{\rm BH} - n_{\rm sph}$ relation by \cite{2016ApJ...821...88S}}\\
      \tablefoottext{g}{In units of $M_{\odot}$, using the $M_{\rm BH} - M_{\rm sph}$ relation by \cite{2016ApJ...817...21S}}\\
      \tablefoottext{h}{In units of $M_{\odot}$, using the $M_{\rm BH} - PA$ relation by  \cite{Davissubmitted}}\\

      }
\label{sec-spec_x_III}
\end{sidewaystable*}


\subsection{BH -- Spheroid scaling relations}

Our first estimate of the BH masses is based on the $M_{\rm BH} - M_{\rm sph}$ scaling relation for spiral galaxies in \cite{2016ApJ...817...21S}, where $M_{\rm sph}$ is the spheroid absolute magnitude at 3.6\,$\mu$m. The $I$-band spheroid magnitudes of the three galaxies observed with $HST$ ($F814W$) were converted to 3.6\,$\mu$m using a colour of ($F814W$ - 3.6\,$\mu$m) $ = 3.53\pm0.03$ (see Ciambur et al. in prep.)

\cite{2016MNRAS.460.3119S} have shown that the $M_{\rm BH} - L_{\rm sph}$ relation for massive early-type galaxies is biased, such that it over-estimates the BH mass in galaxies for which no direct BH mass measurement is available. The bias arose from the need to spatially resolve the BH's sphere of influence when constructing samples with directly measured BH masses. {\it If} such a bias exists among the late-type galaxies, then it would be necessary to reduce the predicted masses in column 11 of Table \ref{sec-spec_x_III}, perhaps by a factor of 3 to 5.

We calculated a second estimate of the central black hole masses from the major axis S\'ersic indices of the spheroid components, $n_{\rm sph}$, based on the $M_{\rm BH} - n_{\rm sph}$ relation for all galaxies, reported in \cite{2016ApJ...821...88S}. The author does provide separate $M_{\rm BH} - n_{\rm sph}$ relations for spiral and early-type galaxies, but due to the small difference she prefers a single, better constrained relation obtained from the entire sample of galaxies.

\subsection{BH -- $PA$ scaling relation}

We use the latest  $M_{\rm BH} - PA$ relation, derived by \cite{Davissubmitted} using a sample of 43 directly measured SMBH masses in spiral galaxies. The linear fit is:
\begin{equation}
\log({M_{\rm BH}/M_{\sun}}) = (9.01\pm0.08) - (0.130\pm0.006)|PA|,
\end{equation}
with intrinsic scatter $\epsilon=0.22^{+0.04}_{-0.05}$ dex in $\log({M_{\rm BH}/M_{\sun}})$ and a total absolute scatter of 0.44 dex. The quality of the fit can be described with a correlation coefficient of -0.85 and a $p$-value of $7.73\times10^{-13}$. For each galaxy, the $PA$ and its corresponding BH mass estimation are listed in Table~\ref{sec-spec_x_III} (columns 5 and 12 respectively).

  \subsection{Fundamental plane of black hole activity}
  \label{FP}
 We use the FP-BH by \cite{2012MNRAS.419..267P} to estimate the mass of the central BHs of all sources in our list. The FP-BH is described by the following regression: ${\rm log}L_{\rm X}=(1.45\pm0.04){\rm log}L_{\rm R}-(0.88\pm0.06){\rm log}M_{\rm BH} -6.07\pm1.10$, where $L_{\rm R}$ and $L_{\rm X}$ have units of erg/s and $M_{\rm BH}$ is in solar masses.
Intrinsic scatter in the FP-BH is smallest when including AGN in accretion states that are similar to {\it hard-state}  (a state characterized by sub-Eddington accretion and hard, power-law-shaped spectral energy distributions) X-ray binaries (XRBs) \citep[e.g.,][]{2001ApJ...548L...9M, 2003MNRAS.345L..19M, 2006MNRAS.372.1366K}. There is a well observed global radio/X-ray correlation in XRBs, in the low-accretion, non-thermal regime \citep[e.g.,][]{2003MNRAS.344...60G}, and the FP-BH, is, in essence, an expansion of this correlation in the higher mass regime. 
Indeed, the study of unobscured AGN suggests an accretion state division that is similar to XRBs \citep[e.g.,][]{1999MNRAS.304..160J,2000ApJ...530L..65N,2001MNRAS.327..739G,2002ApJ...564...86B,2009MNRAS.396.1929H,2011ApJ...733...60T,2012ApJ...745L..27P}
All sources in our sample are accreting at sub-Eddington rates, and with the exception of NGC~3507 display the characteristic power law spectra, usually associated with advection-dominated accretion flows. We can, therefore,  consider their AGN as more massive analogs to {\it hard-state} XRBs, and use the FP-BH to estimate the mass of their central BH. We also use the FP-BH to calculate the mass of the central BH in NGC~3507, the spectrum of which cannot be modeled using a power law. We point out the caveats of this estimation in section~\ref{discussion}.
All mass estimates are presented in Table~\ref{sec-spec_x_III} along with the radio and X-ray luminosities of each source and the mass estimates derived using the $M_{\rm BH} - L_{\rm sph}$ and $M_{\rm BH} - n_{\rm sph}$ scaling relations. In sources with multiple X-ray observations, and since all fit parameters (and flux estimations) were consistent in the 90\% confidence range, the mean value of the luminosity of each source was used. In the case of NGC~4314, we only considered the \cxo derived luminosities, taking advantage of the high imaging resolution of \cxo, in order to avoid contamination from the obvious, extended, extra-nuclear X-ray emission (see details for NGC~4314 in section~\ref{discussion}). In sources for which radio emission was not detected, a 1$\sigma$ upper limit for the mass of the central BH is estimated.

Mass estimations from the  $M_{\rm BH} - n_{\rm sph}$,  $M_{\rm BH} - PA$ scaling relations and the FP-BH, are plotted in Figure~\ref{fig:dan}, against the spheroid stellar mass ($M_{\rm \star sph}$). Included in the plot are the data points used by \cite{2016ApJ...817...21S}, along with their regressions.

\section{Discussion}
 \label{discussion}

\begin{figure*}
\includegraphics[width=1.0\textwidth]{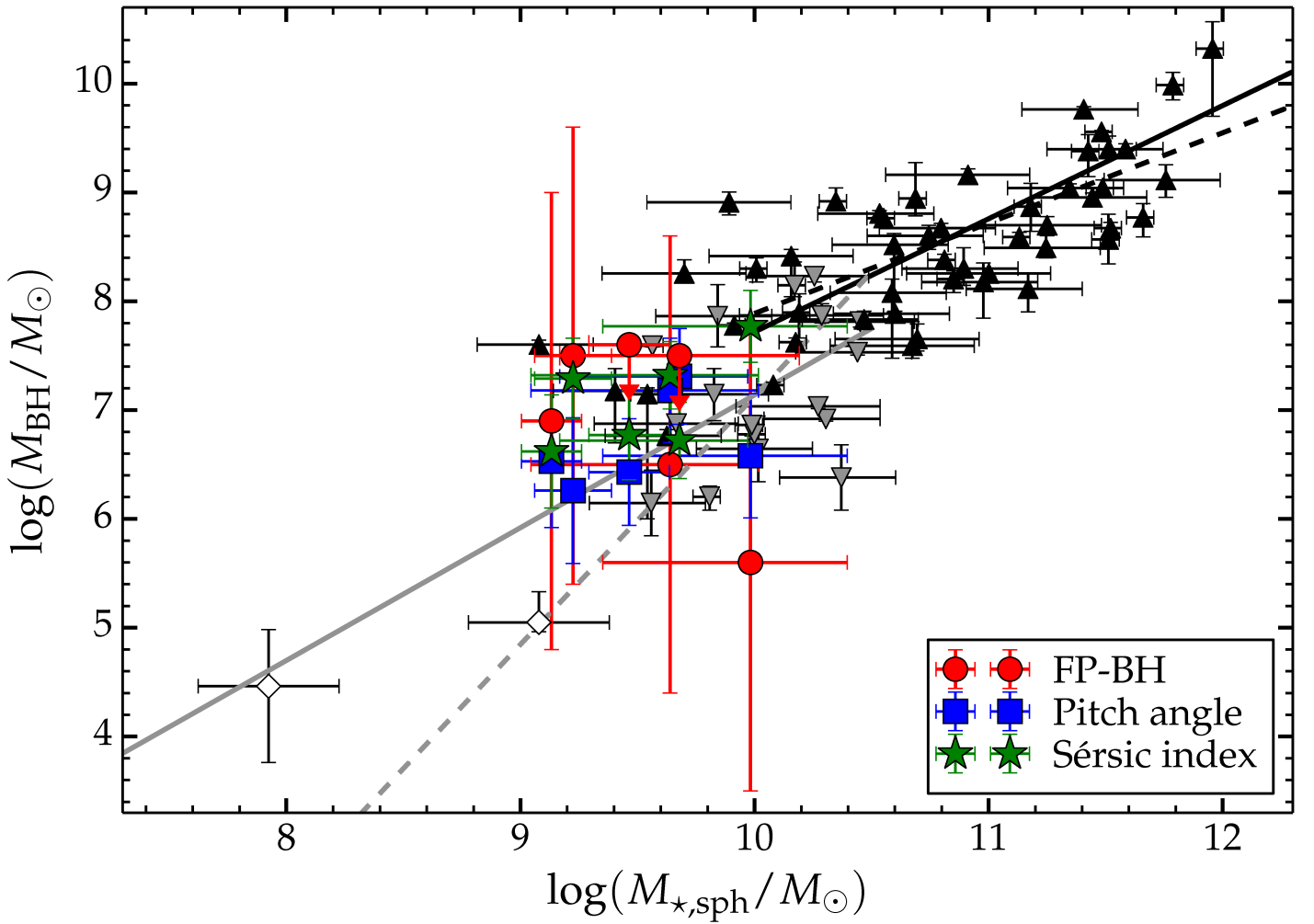}
\label{fig:dan}
\caption{ {\it Mass estimates of the central BH of seven LLAGN (colored points), using our multiple method approach vs. spheroid stellar mass ($M_{\rm \star sph}$).}  Our plot points have been plotted on top of the  scaling relations from \cite{2016ApJ...817...21S} (solid lines: Y on X-axis regression, dashed lines: symmetric regression), the data that the authors used to obtain them as triangles with error bars (black/gray for ellipticals/spirals), the empty diamonds are two other IMBH candidates (LEDA87300: e.g., \citealt{2015ApJ...809L..14B,2016ApJ...818..172G} and Pox52: e.g., \citealt{2004ApJ...607...90B,2008ApJ...686..892T,2016arXiv160708620C}), and the colored data points are our results. The green stars come from the S\'ersic index, the blue squares from pitch angle and the red circles from the FP-BH. The BH masses derived from the bulge luminosity were not plotted, so as not to overcrowd the plot and since the spheroid luminosity ($L_{\rm sph}$) is more or less directly related to the spheroid stellar mass ($M_{\rm \star sph}$) via the mass-to-light ratio. }

\end{figure*}

We have combined the latest $M_{\rm BH} - L_{\rm sph}$, $M_{\rm BH} - n_{\rm sph}$ , and  $M_{\rm BH} - PA$ scaling relations and used the most recent, state-of-the-art tools in IR imaging analysis, to estimate the mass of the central BH in seven LLAGN.
We find that, while all the galaxies in our sample harbor lower-mass-range BHs ($\log{M_{\rm BH}/M_{\odot}}{\approx}6.5$ on average), their mass is significantly higher than for IMBHs ($\leq 10^5\,{\rm M}_\odot$). This result is supported by all four methods used and applies to all sources. All mass estimations from all methods are consistent within the 1\,$\sigma$ error bars.

The $M_{\rm BH}$ estimations in this work, derived using the \citeauthor{2016ApJ...817...21S} $M_{\rm BH} - L_{\rm sph}$ relation, are significantly and consistently higher than in GS13. This is due to several factors that make our analysis different to that of GS13. 

GS13 estimated the BH masses based on the bulge magnitudes computed by D\&D06, after correcting for dust absorption and the (newly discovered at the time) ``near-quadratic'' scaling relation for S\'ersic galaxies. However,  D\&D06 obtained these magnitudes from simple bulge/disc decompositions of ground-based ({\it 2MASS}) imaging. As we have noted in \ref{opt-IR}, most of our galaxies (NGC~3185\footnote{\cite{2013MNRAS.431.3060E}}, 3198\footnote{\cite{2005MNRAS.360.1201H}}, 2486\footnote{\cite{2015AJ....149....1Z}}, 3507\footnote{\cite{2013MNRAS.431.3060E}}, 4314\footnote{\cite{1992AJ....103..757B}}) are not, in fact, simple bulge/disc systems, but are clearly barred. This is reflected in their morphological classification (see Table~\ref{sec-spec_x_III}), and is easily discernible in their images and surface brightness profiles. Furthermore, the presence of AGN is usually associated with a point source component at the center (i.e., a component shaped like the PSF in the surface brightness profile; see the green component in Figure~2, for example). The surface brightness profile is a superposition of all these components, and in order to correctly extract the properties of the S\'ersic spheroid, each component needs to be modeled. As such, in the present work we analyzed superior, space-based, higher-resolution data ({\it HST} and {\it Spitzer}), and performed much more detailed decompositions than D\&D06. Our spheroid magnitudes are consequently more accurate than those obtained in D\&D06, and used in GS13.

Furthermore, GS13 divided their galaxy sample into core-S\'ersic galaxies, which follow a near-linear ($M_{\rm BH}-L_{\rm sph}$) scaling relation, and S\'ersic galaxies, where the relation becomes near-quadratic. They estimated the BH masses from the latter, specifically from the $symmetric$ relation, that is, the bisector between a regression along the $y$-axis ($M_{\rm BH}$) and a regression along the $x-$axis (bulge luminosity). On the other hand, \citeauthor{2016ApJ...817...21S} (whose scaling relations are employed in the present work) divided their galaxy sample into early-type galaxies, which follow a near-linear scaling relation, and spiral galaxies, which follow a near-quadratic relation. The symmetric, near-quadratic relation of \cite{2016ApJ...817...21S}, though an updated version from GS13, nevertheless agrees very well with the latter \citep[see][their Figure~1]{2016ApJ...818..172G}. However, in predicting the BH masses, we did not employ the symmetric relation but rather that obtained by regressing along the $M_{\rm BH}$ axis ($y$ on $x$, or Y|X regression), because this minimizes the uncertainties in the predicted quantity of interest \citep[see Section V in][]{1990ApJ...364..104I}.  Due to the considerable scatter and relatively low number of data points for spiral galaxies, the Y|X regression is noticeably shallower than the symmetric relation\footnote{see the solid and dashed grey lines in Figure~\ref{fig:dan}, for the ($M_{\rm BH}$ -- $M_{\rm \star sph}$) relation}, and this accounts in part for the systematically higher BH masses we obtain, compared to GS13.

For the FP-BH calculations, we used  proprietary radio observations and archival X-ray observations. The obvious caveat of this scheme is the fact that the radio and X-ray observations were not contemporaneous. All radio observations were carried out in 2015 and as a result, their chronological separation from the archival X-ray observations ranges from several months to more than a decade. This time gap may increase the risk of comparing X-ray and radio luminosities of a given source during different AGN ``states'', and thus over- or under-estimating the mass of the central BH. However, while the launching and quenching of radio jets and the transition between soft, accretion-dominated and hard, non-thermal states in XRBs takes place in a human timescale of months or years, in AGN, similar episodes are expected to scale up with the mass of their central BH, bringing their expected transition timescales to $10^3$ years or higher \citep[e.g.,][and references therein]{2015ASSL..414...45T}. The relative stability of AGN luminosity, and particularly of LLAGN, is strongly  supported by observations \citep[e.g.,][]{2012A&A...544A..80G}. It is also hinted at by our own analysis of multiple X-ray observations of NGC~628 and NGC~4470, spanning more than a decade, during which their spectral shape and X-ray flux remain consistent. 

An apparent increase (of a factor of $\sim$4) in the X-ray flux of NGC~4314, between \cxo and \xmm observations, is the result of contamination of the nuclear X-ray emission in the \xmm observation by an apparently extended X-ray source with a radius of $\sim$a few kpc, centered around the galactic center.
The estimated source flux from the \xmm observation, where we extracted a spectrum from a 30\arcsec\ circle (instead of 3\arcsec\ in \cxo),  is $\sim$3.6 times higher than the source flux in the \cxo observation. Indeed, when extracting a spectrum from a 30\arcsec\ circle in the \cxo observation, the  source flux is the same as in the \xmm observation. While the extended emission contributes significantly to the AGN emission, it does not rule out its presence. Most of the X-ray emission appears to originate from the central engine. This is why a hundred-fold increase in the extraction area only results in a less than four times increase of the total flux. The seemingly diffuse component could be unresolved emission from a fairly large population of HMXBs. This hypothesis is further supported by the detection of a ring-like, circumnuclear  radio emission. Also detected by \cite{1991A&A...244..257G}, this emission is thought to originate from HII regions, ionized by the presence of massive stars, together with non-thermal emission from relativistic particles produced during explosions of these massive stars. 

Most sources in our sample appear to be in a state that is equivalent to the XRB {\it hard state}. Namely, they show no evidence of thermal emission and their spectra are well described by power-law distributions. As discussed in section~\ref{FP}, using {\it hard state} sources and the FP-BH to estimate their $M_{\rm BH}$ strengthens the validity of our results. However, NGC~3507 does not follow this pattern. The source emission is consistent with ionized plasma emission  and 98\% of registered photons have energies below 3\,keV. The X-ray continuum does not follow a power-law distribution. What appears as nuclear X-ray emission could be due to a fairly large concentration of supernova remnants (SNRs) in a starburst galaxy. Starburst galaxies are known to host large groups of SNRs  \citep[e.g.,][]{2001ApJ...558L..27C} whose unresolved emission would be more consistent with the observed soft X-ray emission than the emission from a population of HMXBs, which would be considerably harder. Indeed, the presence of a young stellar population in the central region of NGC~3507 \citep{2004ApJ...605..127G} increases the likelihood of the starburst hypothesis. On the other hand, ${\rm O_{III}/H\alpha}$ and ${\rm N_{II}/H\alpha}$ ratios \citep{2015ApJS..218...10V} place NGC~3507 in the AGN region of the BPT diagram and we cannot rule out an active nucleus. However, we have serious doubts over this classification. Therefore, our FP-BH-based estimation of the $M_{\rm BH}$ in this source, the largest mass estimate in our sample using this method, should be considered with caution.

In sources NGC~628, 4314, and 4470 we did not detect a radio counterpart. In these cases, we calculated a 1$\sigma$ upper limit on the mass of the central BH.
The value of the upper limit is determined by the uncertainties in the FP-BH coefficients, which result in substantially large error bars for the mass estimates ($ d({\rm log}M_{\rm BH}//M_{\odot}){\sim}2$). It is important to stress that mass estimations using the FP-BH suffer from a relatively large uncertainty, regardless of the accuracy of the flux and distance estimations. Consequently, upper limits derived from the FP-BH are considerably large. Nevertheless, longer radio observations may achieve detection of the radio counterpart and therefore provide a mass estimation. 
Providing an FP-BH derived upper limit for the $M_{\rm BH}$ implies an underlying assumption that these three sources have central BHs that are accreting material and are also producing relativistic jets, whose radio emission was not detected due to the short exposure of our proprietary observations. Nevertheless, there are strong indications supporting this assumption. All three sources are known LLAGN, exhibiting significant H$\alpha$ \citep{1997ApJS..112..315H} and compact X-ray emission (this work) from their nuclear region. Furthermore, NGC~628 and 4314 have both been detected in the radio in the NRAO VLA Sky Survey \citep[NVSS,][]{1998AJ....115.1693C}. However, the NVSS is a relatively low-resolution survey, and, as such, the registered radio emission cannot be traced back to a potential relativistic jet.

It is evident that, when using the FP-BH to determine the BH mass in LLAGN, the highest angular resolution in X-ray and radio observations is desired, as well as adequate observing time. It is important to stress the fact that our proprietary radio observations had short durations and the archival X-ray observations were often short and the source of interest was considerably off-axis. Future, proprietary, contemporaneous \cxo and {\it VLA} or {\it VLBA} observations with sufficiently long duration will provide far superior results. Particularly, adequately long {\it VLBA} observations will not only increase the chance of detection, but also reduce extra-nuclear contamination of the AGN emission, thanks to its high angular resolution. To this end, next generation X-ray telescopes such as {\it Athena} \citep{2013arXiv1306.2307N, 2013arXiv1308.6784B} will offer the necessary combination of angular resolution and flux sensitivity, to not only further improve mass estimations, but help to better constrain the FP-BH itself, reducing the uncertainties of its coefficients. 

The combination of the FP-BH calculations with the results of the BH-(galactic properties) relations provides an invaluable consistency check, since the four methods are largely independent. Our results demonstrate that the FP-BH upholds the results of the $M_{\rm BH}$ scaling relations. More than that, all four relations used in our calculations produced consistent results. This consistency verification, which was one of the two main objectives of this study, is a reassuring result that not only increases the robustness of our estimations, but also the validity of the employed methods. Furthermore, our approach does not require the use of velocity dispersions, and, as such, avoids the issue of offset barred-galaxies/pseudobulges (most galaxies in our sample are barred and have strong indications of pseudobulges) in the $M_{{\rm BH}}$--$\sigma$ diagram, discovered by \cite{2008ApJ...680..143G} and \cite{2008MNRAS.386.2242H}. Nevertheless, we intend to use our multiple method approach to address the issue of the applicability of the  $M_{{\rm BH}}$--$\sigma$  relation in the low mass regime and in barred/pseudobulge galaxies (see also discussion in \citealt{2013ARA&A..51..511K} and \citealt{2016ASSL..418..263G}), in a separate publication.

Due to their (relatively) low mass, the gravitational sphere of influence of IMBHs cannot be spatially resolved with currently available observatories. The use of indirect methods in the form of known scaling relations is required in order to probe the low-mass regime. However, since all scaling relations and the FP-BH have been primarily calibrated using high-mass BHs, for which direct measurements were available, a consistency test in the low-mass regime was an important step of our campaign. To our knowledge, this is the first time four different methods have been used to simultaneously estimate the mass of the same BHs. Our multiple-method approach helps to ensure the validity of future findings, in case of outliers from any one relation. Nevertheless, despite these initial encouraging results, we must stress that this is a pilot study that involved only seven sources. Furthermore, all relations used have weak points, such as the large intrinsic scatter of the FP-BH, the different regression slopes of the $M_{\rm BH} - L_{\rm sph}$ relation (a consequence of the relatively small available galaxy sample in the low-mass end), and the lack of a robust physical interpretation of the $M_{\rm BH} - PA$ relation. Additionally, the $M_{\rm BH} - PA$ estimations may suffer from wavelength-dependent estimations of the pitch angle (see e.g., \citealt{1998A&A...336..840G}; \citealt{2012ApJ...744...92M}; \citealt{2014ApJ...793L..19M}, but also \citealt{2008ApJ...678L..93S}; \citealt{2012ApJS..199...33D}).

Encouraged by the favorable results of this study, we aim to use our multi-relation approach to add many more low-mass/low-luminosity candidates to this sample.
Furthermore, as this project evolves, we will use the methodology presented here to re-estimate the mass of stronger LLAGN-IMBH candidates in the GS13 list. It is important to note that the sources in our sample were on the high-mass end of the GS13 list. The sources analyzed in this work were chosen primarily because they had relatively high X-ray luminosities (however, still in the LLAGN regime), and had all been detected in archival X-ray observations. There are thirteen sources in the GS13 sample that have estimated\footnote{Using the GS13 $M_{\rm BH} - L_{\rm sph}$ relation} masses of less than $\sim10^4\,{\rm M}_\odot$  and, when using the \cite{2006AJ....131.1236D}  bulge magnitudes and taking into account our new robust $M_{{\rm BH}}$--$L$ relation \citep{2016ApJ...817...21S} and our improved imaging analysis techniques, are still expected to lie  within the IMBH range. For the following steps of this campaign, we intend to obtain X-ray and radio proprietary observations of all thirteen candidates, and with the addition of already available archival IR and optical observations, we will employ our multiple-method approach to re-calculate the mass of their central BH.

\section{Conclusion}
\label{conclusion}   
We have used the FP-BH \citep[re-calibrated by][]{2012MNRAS.419..267P}, the recently revised $M_{\rm BH}$-spheroid relations by \cite{2016ApJ...821...88S} and \cite{2016ApJ...817...21S}, and the  $M_{\rm BH} - PA$ relation by  \cite{Davissubmitted}, to re-estimate the masses of seven LLAGN, which previous calculations placed in the intermediate mass regime. We find that although the central BHs in all seven galaxies have relatively low masses, they are not IMBHs. 
Furthermore, we demonstrate that the combination of radio and X-ray observations and the FP-BH, with IR and optical observations and the latest $M_{\rm BH}$-(galactic properties) relations produces consistent and robust results. This is the first time that the consistency of $M_{\rm BH}$-spheroid relations and the FP-BH have been investigated in the low-mass regime.
The consistency between our predictions strengthens our confidence in the legitimacy of this approach and the techniques it involves. Nevertheless, to ensure optimal future results, the highest angular resolution is desired in all wavelengths, along with sufficient radio and X-ray observing times to ensure source detection. The present work is the start of an ongoing project in search of IMBH candidates in LLAGN.

\section{Acknowledgements}

This research has made use of the NASA/IPAC {\it Extragalactic Database} (NED), and the NASA/ IPAC {\it Infrared Science Archive} (IRSA), which are operated by the Jet Propulsion Laboratory, California Institute of Technology, under contract with the National Aeronautics and Space Administration. Part of this work is based on observations made with the NASA/ESA {\it Hubble Space Telescope}, and obtained from the {\it Hubble Legacy Archive}, which is a collaboration between the {\it Space Telescope Science Institute} (STScI/NASA), the {\it Space Telescope European Coordinating Facility} (ST-ECF/ESA) and the {\it Canadian Astronomy Data Centre} (CADC/NRC/CSA).

BCC acknowledges support from the Embassy of France in Australia (through the Scientific Mobility Program) and from the Australian Astronomical Society.

\bibliography{general}

\end{document}